\documentclass{osa-article}
\journal{oe} 
\articletype{Research Article}

\usepackage[title]{appendix}
\usepackage[utf8]{inputenc}
\usepackage[english]{babel}
\usepackage{physics,mathtools,gensymb}
\usepackage{float,subfig}
\usepackage[section]{placeins}
\usepackage[capitalize]{cleveref}
\creflabelformat{equation}{#2\textup{#1}#3}
\newcommand\F{\mathcal{F}}

\begin{document}

\title{Cavity-enhanced absorption sensing with robust sideband locking}

\author{Fernanda C. Rodrigues-Machado,\authormark{1} Pauline Pestre,\authormark{1} Vincent Dumont,\authormark{1} Erika Janitz,\authormark{1,2} Liam Scanlon,\authormark{1} Shirin A. Enger,\authormark{3,4,5} Lilian Childress,\authormark{1} and Jack Sankey\authormark{1,*} }

\address{\authormark{1} Department of Physics, McGill University, 3600 Rue University, Montreal, QC H3A 2T8,
Canada\\
         \authormark{2} Department of Physics, ETH Zurich, Otto-Stern-Weg 1, 8093 Zurich, Switzerland\\
         \authormark{3} Medical Physics Unit, Department of Oncology, Faculty of Medicine, McGill University, Montreal, QC H4A 3J1, Canada\\
         \authormark{4} Research Institute of the McGill University Health Centre, Montreal, QC H4A 3J1, Canada\\ 
         \authormark{5} Lady Davis Institute for Medical Research, Jewish General Hospital, Montreal, QC H3T 1E2, Canada}

\email{\authormark{*} jack.sankey@mcgill.ca }

\begin{abstract}
We present a simple, continuous, cavity-enhanced optical absorption measurement technique based on high-bandwidth Pound-Drever-Hall (PDH) sideband locking. The technique provides a resonant amplitude quadrature readout that can be mapped onto the cavity’s internal loss rate, and is naturally compatible with weak probe beams. With a proof-of-concept 5-cm-long Fabry-Perot cavity, we measure an absorption sensitivity of $6.6\times 10^{-11}~\text{cm}^{-1}/\sqrt{\text{Hz}}$ at 100 kHz (roughly the cavity bandwidth) and $\sim 10^{-10}~\text{cm}^{-1}/\sqrt{\text{Hz}}$ from 30 kHz to 1 MHz, with 38~$\upmu$W collected from the cavity's circulating power.
\end{abstract}

\section{Introduction and motivation}

Cavity-enhanced sensing is a powerful tool for measuring weak optical absorption signals, with applications ranging from molecular spectroscopy to biomolecular sensors~\cite{romanini2014}. By confining light within a cavity, the effective optical depth of its interaction with any intra-cavity material is enhanced by $2\mathcal{F}/\pi$~\cite{ye2003}, where $\mathcal{F}$ is the finesse of the cavity, making it possible to detect very weak changes in absorption with a compact apparatus.  A wide variety of application-specific absorption sensing methods exist, usually based on detection of transmitted light \cite{hodgkinson2013,goldenstein2017,cossel2017,hu2021} or measurement of the cavity lifetime~\cite{maity2020}.  Other sensing modalities employ heterodyne detection to boost weak signals above technical noise, such as the NICE-OHMS scheme~\cite{ye1998, ma1999}, which allows for shot-noise-limited readout of narrow absorption features. 

To continuously exploit the $2\mathcal{F}/\pi$ enhancement, the probe light must be locked on resonance with the cavity. One simple solution is to employ a recently developed Pound-Drever-Hall~\cite{black2001} (PDH)-based locking scheme that achieves a robust lock with MHz-bandwidth (delay-limited) control over a phase-modulation sideband~\cite{reinhardt2017}. This technique is well suited to sensing narrow \emph{and} wide absorption features, especially in situations requiring a weak probe beam (the signal is heterodyne-amplified), a comparatively simple apparatus (one laser, one modulator, and low-cost RF electronics), and / or single-port access (e.g., endoscopy). Under ideal circumstances, this heterodyne reflection signal presents the same shot-noise-limited absorption sensitivity as a transmission measurement with a symmetric cavity of the same finesse and circulating power.

Using a cm-scale, finesse-9,000 Fabry-Perot cavity as a testbed and subtracting some of the laser's classical noise, we measure a sensitivity of $10^{-10}~\text{cm}^{-1}/\sqrt{\text{Hz}}$ for frequencies between 0.03-1~MHz and a minimum sensitivity of ${6.6\times 10^{-11} ~\text{cm}^{-1}/\sqrt{\text{Hz}}}$ at 100 kHz. This measurement is performed with a total of 3.5~mW landing on the cavity, with a sideband coupling 160~$\upmu$W into the cavity, and 38~$\upmu$W of the former cavity light collected by our ``signal'' photodiode (after losses from our diagnostic beam splitters). Without optimizing our choice of laser or modulation electronics, the minimum sensitivity is within a factor of $\sim$10 of the shot noise level, demonstrating a straightforward means of achieving low-noise transient readout with MHz bandwidth (limited by the cavity ringdown rate). 


\section{Measurement Scheme and Fundamental Limits}

\begin{figure}[ht]
    \centering
    \includegraphics[width=0.6\textwidth,trim=0 0 0 0,clip]{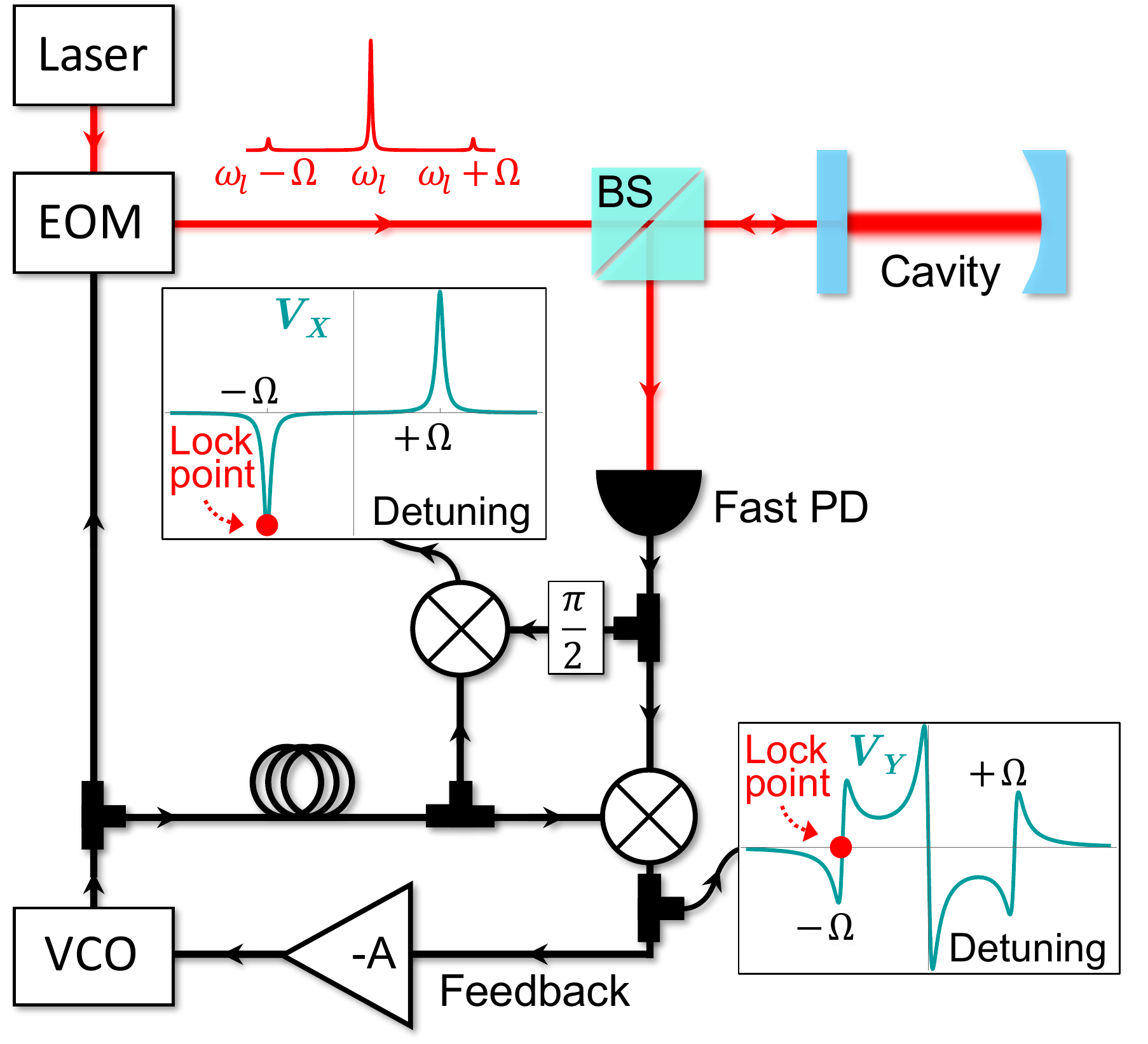}
    \caption{\textbf{Basic measurement scheme.} Laser light at frequency $\omega_l$ is phase-modulated at frequency $\Omega$ by an electro-optical modulator (EOM) driven by a voltage-controlled oscillator (VCO), generating sidebands at ${\omega_l\pm\Omega}$. These beams are reflected from an optical cavity, generating amplitude modulation that is detected by a ``fast'' photodiode (PD). The PD signal and a $\pi/2$-shifted copy are simultaneously demodulated by a delay-matched copy of the original VCO signal, producing the steady-state, detuning-dependent voltages $V_X$ (``amplitude quadrature'', upper inset) and $V_Y$ (``phase quadrature'', lower inset). At the indicated sideband lock point, $V_Y$ provides an error signal to be fed back to the VCO via amplifiers and filters (transfer function $-A$), while $V_X$ records cavity losses. }
    \label{fig:setup}
\end{figure}

To summarize the technique, Fig.~1 shows the employed optical circuit~\cite{reinhardt2017}, which locks a laser sideband to a cavity resonance while monitoring intra-cavity losses. First, continuous-wave laser light at frequency~$\omega_l$ is phase modulated by an electro-optical modulator (EOM) driven by a voltage-controlled oscillator (VCO) at frequency~$\Omega$, thereby creating sidebands at frequencies ${\omega_l \pm \Omega}$. 
When the laser or a sideband is near a cavity resonance, phase modulation is converted to amplitude modulation in the reflected light, which is detected and demodulated with a delay-matched copy of the VCO output. The resulting ``phase'' quadrature $V_Y$ varies linearly with detuning, providing an error signal that is fed back to the VCO.  In contrast, demodulation of a $\pi/2$-shifted copy of the signal yields the ``amplitude'' quadrature $V_X$; this value provides information about intra-cavity losses when a sideband is on resonance with the cavity, as discussed below. Note this scheme exploits the carrier as an inherently path-matched heterodyne local oscillator, which boosts the signals (both $V_X$ and $V_Y$) from the resonant sideband above the detector noise.

To predict the scheme's fundamental sensitivity limit, consider a cavity of length $L$ with front and back mirrors of amplitude reflectivities $-r_1$ and $-r_2$, respectively (assuming real-valued $r_i$ for simplicity) 
and a ``baseline'' power loss fraction $2\delta$ per round trip (from just the mirrors). We model any medium inside the cavity as providing a power absorption coefficient per unit length $\alpha$. In the limit of high reflectivity, low intrinsic loss, and low absorption, the cavity finesse is
\begin{equation}\label{eq:Finesse}
\F\approx \pi/(1- r_1 r_2 + \alpha L + \delta).
\end{equation} 
For an incident phase-modulated laser beam (electric field ${ \propto e^{i\omega_l t + i\beta \sin{\Omega t}} }$), a sideband that is exactly on resonance will experience an amplitude reflection coefficient 
    \begin{align}\label{eq:rres}
    r_{\text{res}} \approx -r_1 + \F r_2 t_1^2/\pi,
    \end{align}
where ${t_1 =\sqrt{ 1- r_1^2 }}$ is the front mirror amplitude transmission (mirror losses are already captured by~$\delta$). Meanwhile, if the carrier and other sideband are sufficiently far detuned from resonance (i.e., in the ``resolved-sideband'' limit, where $\Omega$ is much larger than the cavity ringdown rate $1/\tau$), they will experience a cavity reflectivity~${\approx -r_1}$. In this configuration, the reflected power becomes \cite{black2001, reinhardt2017}
    \begin{equation}    \label{eq:Ppdh}
    P_{R, \mathrm{PDH}} = \underbrace{ (1-\epsilon) P_\mathrm{in} r_1^2+\epsilon P_\mathrm{in} \left(\rho_{\mathrm{c}} r_1^2 + \rho_{\mathrm{sb}}(r_1^2 + r_{\text{res}}^2) \right) }_{\bar{P}} + \underbrace{\epsilon P_\mathrm{in} \left( 2\sqrt{ \rho_{\mathrm{c}} \rho_{\mathrm{sb}} } (r_1^2 + r_1 r_{\text{res}}) \right) }_{P_X}\cos(\Omega t) ,
    \end{equation}
where $\rho_{\mathrm{c}}$ and $\rho_{\mathrm{sb}}$ are the fraction of the input power going into the carrier and each sideband respectively (${\rho_{\mathrm{c}} + 2\rho_{\mathrm{sb}} \approx 1}$ in the weak-modulation limit), and $P_\mathrm{in}$ is the total incident power. We drop terms oscillating at $2\Omega$ (electronically filtered in the experiment), and explicitly account for imperfect mode-matching to the cavity via a power coupling fraction $\epsilon$, where ${\epsilon = 1}$ corresponds to perfect coupling; the first term in \cref{eq:Ppdh} accounts for reflection of uncoupled light. Note this \emph{resonant} expression exhibits an \emph{out-of-phase} beat note of amplitude $P_X$, while the in-phase beat note $\propto \sin(\Omega t)$ is only nonzero when the sideband is detuned from resonance. After demodulation, this beat note produces a measured voltage ${V_X = G P_X}$ with a gain $G$ including all conversion factors associated with the photodetector and mixer circuitry. 

Importantly, $V_X$ depends on $r_{\text{res}}$, which itself depends on the cavity finesse (Eq.~\ref{eq:rres}) and can therefore provide a continuous record of the absorption coefficient $\alpha$ (Eq.~\ref{eq:Finesse}). The sensitivity of such a measurement depends on the variation of $P_X$ with $\alpha$ ($|dP_X/d\alpha|$) as well as how much noise is present. When detection is limited by the shot noise $\langle S_P\rangle = \sqrt{2 \bar{P} e/\eta}$ on the total average power $\bar{P}$ reaching the detector (where $e$ is the electron charge and $\eta$ is the detector responsivity), the sensitivity becomes
    \begin{equation} \label{eq:sa}
    \langle S_\alpha\rangle = \frac{\sqrt{2}\langle S_P\rangle} {|d P_X/d\alpha|}
     = \frac{\pi^2} 
     { \F^2 L t_1^2}
     \frac{R} {\epsilon \sqrt{\bar{P} \rho_{\mathrm{c}} \rho_{\mathrm{sb}}} }
     \sqrt{ \frac{e}{\eta} },
    \end{equation}
where $R = 1+ \epsilon \left(-1 + \rho_{\mathrm{c}} r_1^2 + \rho_{\mathrm{sb}}(r_1^2+r_{\text{res}}^2)\right) \approx 1$ is the average reflected power fraction, and the factor of $\sqrt{2}$ arises from noise necessarily added by the demodulation process. Note this expression assumes that absorption changes slowly compared to the cavity lifetime; the sensitivity to absorption falls of at higher frequencies, as discussed below.

\begin{figure}
    \centering
    \includegraphics[width=0.43\textwidth,trim=0 0 0 0,clip]{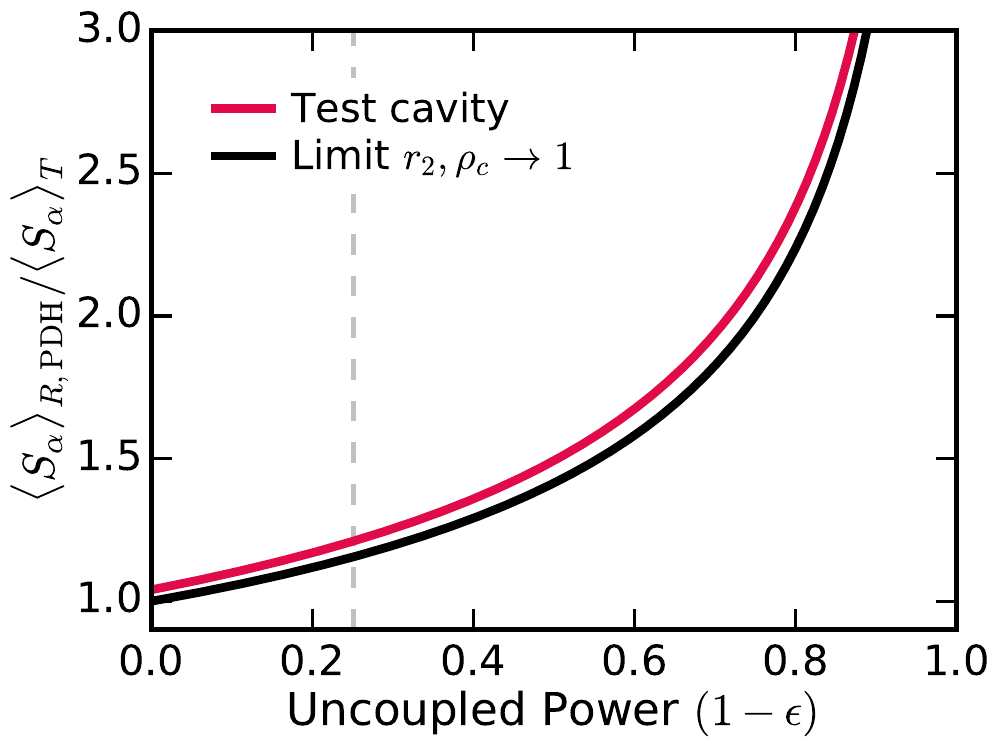}
    \caption{\textbf{
    Shot-noise-limited sensitivity of sideband-locking technique relative to that of a transmission measurement through a symmetric cavity.} 
    The black curve shows the relative sensitivity $1/\sqrt{\epsilon \rho_{\mathrm{c}}}$ versus uncoupled power fraction $1-\epsilon$ for an ideal cavity with $r_2,\ \rho_{\mathrm{c}}\rightarrow 1$ and $\rho_{\mathrm{sb}} \to 0$. The red curve corresponds to our chosen (commercially available) high-finesse mirrors (${r_1=0.999796}$, ${r_2=0.9999974}$), loss parameter ($\delta=143$~ppm) and modulation strength (${\rho_{\mathrm{c}}=0.8831}$, ${\rho_{\mathrm{sb}}=0.0590}$). The gray dashed line shows the coupling level for our system, aligned with free-space optics.}
    \label{fig:sensitivityR2T}
\end{figure}

The black line in Fig.~\ref{fig:sensitivityR2T} shows how $\langle S_\alpha\rangle$ varies with the mode mismatch ${1-\epsilon}$ for an ideal ``single-sided'' cavity (${r_2\rightarrow 1}$) and weak modulation (${\rho_{\mathrm{c}}\rightarrow 1}$), provided we collect all the reflected light. When perfectly coupled ($1-\epsilon=0$), the sensitivity is identical to that of the usual transmission techniques employing ``symmetric'' cavities (${r_1=r_2}$) of the same finesse and circulating power. Note that this sensitivity is a factor of $\sqrt{2}$ worse than can be achieved by monitoring the reflected power alone, or a transmission scheme with a reversed single-sided cavity, but avoids the low-power locking challenges and high input power requirements associated with those approaches, respectively. We also note that (perhaps counter-intuitively) our scheme does work with symmetric cavities despite the absence of reflected power on resonance, though at another penalty of $\sqrt{2}$ in sensitivity. Imperfect cavity coupling (${\epsilon<1}$) adds additional uncoupled power to the reflection photodiode, thereby increasing the shot noise. However, this penalty is a slowly increasing function in the range of ``typical'' values (dashed line). For reference, the red line shows the shot noise limit for the commercially available mirrors discussed below. 

\section{Experimental Setup}\label{sec:setup}

We experimentally test the technique using a 5-cm-long, finesse $\sim$~9,000 cavity probed by a 1550~nm continuous-wave laser beam, phase-modulated at $\Omega/2\pi \sim 1$~GHz. Before the optical cavity, half the input beam is picked off by a non-polarizing beamsplitter to monitor $P_\mathrm{in}$, thereby allowing us to account for power drift and eliminate some classical laser noise (see Sec.~\ref{sec:absorption-sensitivity}). Light reflected from the cavity is detected via the other port of this beamsplitter, where we measure both high- and low-frequency reflection signals; lacking a fast photodiode capable of detecting down to DC, we split the reflected light again so that half of the power lands on a DC-coupled ($\sim 150$~MHz bandwidth) ``diagnostic'' photodiode and the other half lands on an AC-coupled $\sim2$~GHz bandwidth ``signal'' photodiode. Another DC-coupled detector is used to monitor the transmitted signal. 

\begin{figure} 
    \centering
    \includegraphics[width=0.99\textwidth,trim=0 0 0 0,clip]{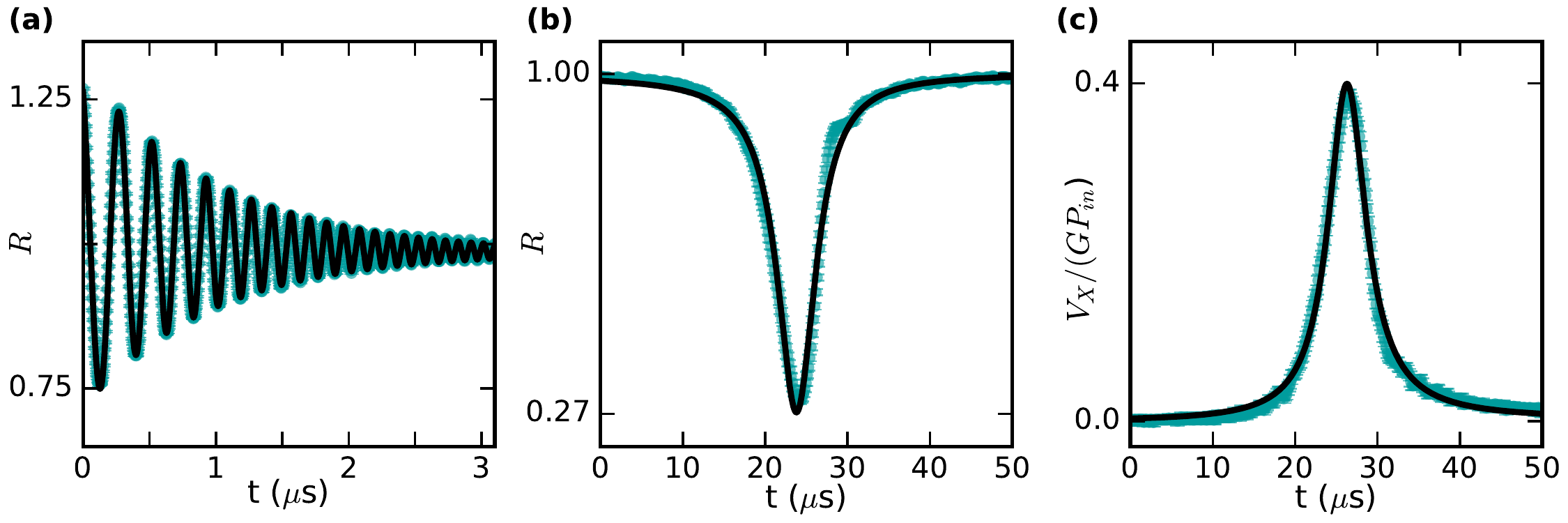}
    \caption{\textbf{Empty cavity characterization with simultaneous fits.} We extract the cavity lifetime~$\tau$, demodulation gain $G$, power coupling $\epsilon$, back mirror reflectivity $r_2$, and carrier (sideband) power fraction $\rho_{\mathrm{c}}$ ($\rho_{\mathrm{sb}}$) by simultaneously fitting six cavity length sweep measurements, examples of which are shown: (a)~swept ring-down spectroscopy; (b)~cavity reflection $R$ without phase modulation; (c)
   ~demodulated PDH beat signal $V_X$ near the lower sideband (transmission $T$  and modulated reflection of carrier and sidebands not shown). Measurements are normalized by off-resonant values in (a) and (b), and $GP_\mathrm{in}$ in (c).
    }
    \label{fig3:fits} 
\end{figure}

To estimate the cavity parameters, we simultaneously fit data from six measurements, three of which are shown in Fig.~\ref{fig3:fits} (see \cref{app:params} for more details). The six independent parameters are: (1)~the empty cavity power lifetime $\tau$, (2)~the coupled power fraction $\epsilon$, (3)~the back mirror reflectivity $r_2$, (4-5)~the carrier and sideband power fractions $\rho_{\mathrm{c}}$ and $\rho_{\mathrm{sb}}$, and (6)~the conversion gain factor $G$.
We first measure the swept-cavity ringdown time \cite{he2002} on the DC-coupled photodetector by rapidly sweeping the cavity length via the piezo-mounted back mirror (Fig.~\ref{fig3:fits}(a)). 
Next, we slowly sweep the cavity length to monitor quasi-steady-state reflected and transmitted power fractions $R$ (Fig.~\ref{fig3:fits}(b)) and $T$ without phase modulation. Finally, we turn on phase modulation and measure the reflected signal and $V_X$ (Fig.~\ref{fig3:fits}(c))  near the lower sideband and carrier frequencies. Fitting the dataset simultaneously (black curves) yields ${\tau = 495.9\pm0.2}$~ns, ${\epsilon = 0.749\pm0.005}$, 
${r_2=0.9999974\pm0.0000001}$, 
${\rho_{\mathrm{c}}=0.8831\pm0.0007}$, ${\rho_{\mathrm{sb}}=0.0590\pm0.0004}$ and ${G=-80\pm3}$~V/W for the empty cavity. 
The remaining parameters, ${t_1^2=0.00041 \pm 0.00001}$, and ${L=5.2 \pm 0.1}$~cm are measured directly.

\section{Absorption sensitivity}\label{sec:absorption-sensitivity}

\begin{figure}[ht]
    \centering
    \includegraphics[width=\textwidth,trim=0 0 0 0,clip]
    {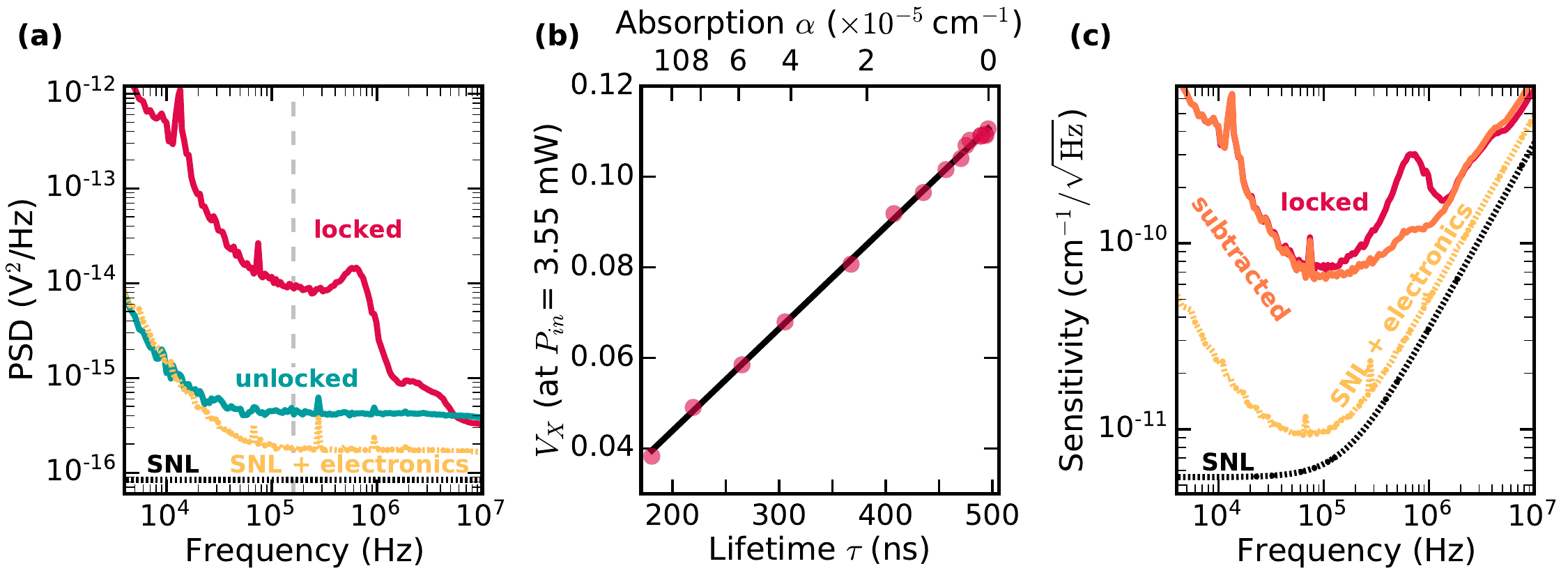}
    \caption{\textbf{Estimation of absorption sensitivity.} 
    (a) Voltage noise power spectral densities (PSDs). Red: noise of $V_X$ with the lower sideband locked to the cavity. Teal: noise while unlocked and off resonance. Yellow: shot-noise limit (SNL) and electronics noise (laser off). Black: SNL for the collected light ($625\ \upmu$W). The vertical dotted line marks our $V_X$ measurement bandwith ($1/(4\pi\tau)=160$~kHz).
    (b) $\tau$-dependence of $V_X$ for the sensitivity estimate outlined in the main text (Eq.~\ref{eq:dVxdalpha}). Black curve is a linear fit (see main text). Top axis shows predicted $\alpha$-dependence of $V_X$ ($\alpha=-\frac{1 - r_1r_2 + \delta}{L} + \frac{n}{c\tau}$).
    (c) Estimated absorption sensitivity for the cavity, including the cavity's dynamical response to absorption changes (Appendix \ref{app:mod_absorption}). Red curve is the raw sensitivity, and the orange curve shows the sensitivity after subtracting (in the time domain) the directly monitored laser noise. The raw measurement noise floor is $\sim 7\times10^{-11}~\mathrm{cm}^{-1}/\sqrt{\mathrm{Hz}}$ at 100 kHz (a factor of 11 above the SNL), and $\sim 2\times10^{-10}~\mathrm{cm}^{-1}/\sqrt{\mathrm{Hz}}$ at 1 MHz (a factor of 6 above the SNL). 
    With subtraction, the noise floor decreases to  a factor of 10 from the SNL at 100 kHz and a factor of 4 at 1 MHz.
    }
    \label{fig:results}
\end{figure}

Once locked to the (lower) sideband, the absorption sensitivity can be estimated from the noise power spectral density (PSD) of $V_X$ (Fig.~\ref{fig:results}(a), red) scaled by $dV_X/d\alpha$ similar to Eq.~\ref{eq:sa}.
In principle, we could predict $dV_X/d\alpha$ from the measured cavity parameters, but since
    \begin{align}   \label{eq:dVxdalpha}
    \frac{dV_X}{d\alpha} = \frac{dV_X }{d\tau} \frac{d\tau}{d\alpha} 
    \qquad \text{and} \qquad 
    \abs{\frac{d\tau}{d\alpha}}\approx \frac{c\tau^2}{n} 
    \end{align}
where $c$ is the speed of light and $n$ is the intra-cavity medium's refractive index ($n\approx 1$ for our air-cavity), measuring $dV_X/d\tau$ provides a more direct and constrained estimate. Figure \ref{fig:results}(b) shows the measured $V_X$ and $\tau$ with the internal cavity loss $\alpha$ varied by partially occluding the cavity mode with a block of anodized aluminum. 
The linear dependence of $V_X$ on $\tau$ is expected as Eqs.~\ref{eq:rres} and \ref{eq:Ppdh} yield ${V_X \propto\F r_1 r_2 t_1^2/\pi}$, and the cavity finesse scales directly with ringdown time $\tau$. The fit slope ${\frac{d V_X}{d\tau} = 225.5\pm0.3}$~mV/$\upmu$s in Fig.~\ref{fig:results}(b) agrees with the value $227\pm16$~mV/$\upmu$s calculated from our system parameters. In \cref{fig:results}(c), the results of \cref{fig:results}(a) and (b) are combined to estimate the absorption sensitivity of our test apparatus. Note the cavity bandwidth ${1/4\pi\tau = 160}$~kHz imposes a low-pass on the absorption signal generated within the cavity (see \cref{app:mod_absorption} for details), which increases the impact of noise at higher frequencies.

For the chosen components, the dominant noise is technical. As shown in Fig.~\ref{fig:results}(a), with the laser off, we  can measure the  noise from our readout electronics, but even with the addition of the theoretical shot noise (gold curve) it lies well below the $V_X$ noise spectra observed when the laser is on and the cavity is locked or unlocked. When unlocked, $V_X$ measures the electronic noise and the ($\sim$flat, classical) laser amplitude noise near the demodulation frequency $\Omega/(2\pi)\sim 1$~GHz away from the carrier (teal curve). In contrast, when a sideband is locked on resonance (red curve), $V_X$ is sensitive to low-frequency amplitude noise from several sources. First, there is the laser's low-frequency classical amplitude noise, combined with the VCO's inherent amplitude noise. In addition, as the feedback loop adjusts the VCO frequency to keep the sideband on resonance, the (frequency-dependent) power of the VCO can change, introducing yet another source of noise. We suspect the latter contributes the majority of noise below our locking bandwidth ($\sim$2 MHz), where the feedback gain is still large enough to modulate the VCO setpoint. Nonetheless, as shown in Fig.~\ref{fig:results}(c), the sensitivity is  $\sim 7.4\times10^{-11}~\mathrm{cm}^{-1}/\sqrt{\mathrm{Hz}}$ at 100~kHz (a factor of 11 above the shot noise limit) for ${38\ \upmu}$W collected from the cavity; at 1~MHz, the measurement noise floor is a factor of 6 above the shot noise limit (SNL).

We find the sensitivity can be further improved by subtracting known sources of classical noise (see Appendix~\ref{app:sub} for details), with removal of classical laser noise yielding the largest improvement. Specifically, the voltage from the photodiode monitoring the input laser power was appropriately scaled and delayed (see Appendix~\ref{app:sub}) before subtracting it from $V_X(t)$ in the time domain, resulting in a noise-reduced signal (orange PSD in Fig.~\ref{fig:results}(c)). This subtraction is effective in eliminating the peak in laser amplitude noise near 500~kHz, and results in a sensitivity of ${\sim 6.6\times10^{-11}~\mathrm{cm}^{-1}/\sqrt{\mathrm{Hz}}}$ at 100~kHz, a factor of 10 above the shot noise limit.   

\section{Outlook}

We demonstrated a simple absorption sensing technique with noise floor ${\sim 10^{-10}~\text{cm}^{-1}/\sqrt{\text{Hz}}}$ from 0.03 to 1~MHz, with a robustly locked, low-power probe. The monitored reflection signal provides an often advantageous single-port alternative to transmission measurements common in such sensing techniques. The presented scheme may also offer opportunities to further improve performance without requiring lower-noise components. For example, an approach in the vein of cavity ringdown could be realized by modulating the laser or sideband amplitude, and the characteristic response time (phase and amplitude shift) of $V_X$ would reveal the cavity lifetime.  

The features of this cavity-enhanced absorption sensing technique are particularly well suited to detection of transient and broadband absorption signals. The sideband light is always on resonance with the cavity, which is essential for transient detection, with the high-bandwidth lock efficiently cancelling detuning noise up to several MHz. While the current testbed system is limited in sensitivity above $\sim 1$~MHz due to the cavity lifetime, smaller cavities (or on-chip resonators) could permit detection of transients at higher frequencies. One example application is radiation dosimetry based on water radiolysis \cite{dosimeter_patent}: when pulses of ionizing radiation deposit energy in water, they produce solvated electrons that exhibit a broad optical absorption feature in the visible and near infrared~\cite{hart1962,herbert2017}. Typically, this signal persists for $\sim$microseconds, and is sufficiently weak that cavity enhancement would be required to observe it within a small volume. With the option to measure in reflection, one could even envision constructing a small-footprint probe with a fiber cavity~\cite{hunger2010} to measure radiation dose with a tissue-equivalent medium at point-like locations.

In summary, we have described how to leverage a high-bandwidth locking scheme to continuously detect intra-cavity absorption. For applications that require a high-bandwidth lock, reflection-based detection, and / or low probe powers, this approach could replace transmission measurements in a range of cavity-enhanced absorption sensing schemes, and could facilitate sensitive detection of transient absorption features.


\begin{appendices}

\section{Extracting System Parameters}\label{app:params}

We perform a set of measurements using the voltages of four photodetectors (the ``pick-off'' from the first beam splitter, cavity transmission, ``diagnostic'' (DC-coupled) slow reflection and fast (AC-coupled) reflection) to extract unknown system parameters. The input mirror's power transmission ${e^{-\delta/2}t_1^2 \approx t_1^2}$ (which gives $r_1=\sqrt{1-t_1^2}$), the cavity length $L$ and the transmission path gain factor $G_T = V_T/P_T$ (where $V_T$ is the transmission diode voltage reading and $P_T$ the power transmitted by the cavity) are directly measured. 
By rapidly sweeping the cavity length by means of a piezo-mounted back mirror, we observe ringdown measurements like the one in Fig.~\ref{fig3:fits}(a) to extract the cavity lifetime $\tau$. In addition, by sweeping the cavity length more slowly, we acquire a set of five measurements: cavity reflection $V_R$ and transmission $V_T$ without phase modulation, cavity reflection with phase modulation around the carrier $V_{Rc}$ and the lower sideband resonance $V_{Rsb}$, and the demodulated PDH amplitude quadrature $V_X$ around the lower sideband resonance. These five curves are then simultaneously fit to extract the remaining parameters (carrier and sideband relative power $\rho_{\mathrm{c}}$ and $\rho_{\mathrm{sb}}$, back mirror amplitude reflectivity $r_2$, power coupling parameter $\epsilon$ and the gain factors in the slow ($G_{\mathrm{slow}}$) and fast ($G$) reflection paths).
To account for incoming power fluctuations, we normalize the voltage sweeps using the readings from the pick-off diode.
The swept ringdown measurements are fit with the function \cite{reinhardt2016}:
	\begin{align} \label{eq:RDfit}
	V_{RD} = a e^{-t/2\tau} \cos{((\omega + bt ) t + c )} + d + g t + h t^2 \ 
	\end{align}
where $\tau$, $a$, $b$, $c$, $d$, $g$, and $h$ are fit parameters ($b$ accounts for acceleration due to mechanical noise, while $g$ and $h$ account for background drifts). The equations for the simultaneous fits are:
	\begin{align}
	\frac{ V_R }{P_\mathrm{in} }   &= G_{\mathrm{slow}} \left( (1-\epsilon)r_1^2 + \epsilon R(\Delta L)\right)  
	\\
	\frac{ V_T }{P_\mathrm{in} }   &= G_T \epsilon T(\Delta L)
	\\
	\frac{ V_{Rc} }{P_\mathrm{in}} &= G_{\mathrm{slow}} \left( (1-\epsilon)r_1^2 + \epsilon ( \rho_{\mathrm{c}} R(\Delta L) + 2\rho_{\mathrm{sb}} r_1^2) \right)	
	\\
	\frac{ V_{Rsb}}{P_\mathrm{in}} &= G_{\mathrm{slow}} \left( (1-\epsilon) r_1^2 + \epsilon (\rho_{\mathrm{c}} r_1^2+\rho_{\mathrm{sb}} r_1^2+ \rho_{\mathrm{sb}} R(\Delta L)) \right)
	\\
	\frac{ V_X }{P_\mathrm{in} }   &= 2 G \epsilon \sqrt{\rho_{\mathrm{c}} \rho_{\mathrm{sb}}}\; r_1 \left( r_1+\Re{r(\Delta L)} \right)
	\end{align}
with the cavity fractional amplitude reflection $r$, fractional power reflection $R$, and transmission $T$ given by the high finesse approximation expressions:
	\begin{align}\label{eq:highF}
	r(\Delta L) \approx \frac{ r_2 t_1^2 } { \frac{\pi}{\F} - 2ik_0\Delta Ln } - r_1
	\\
	R(\Delta L) \approx \frac{ r_2 t_1^2 (r_2 t_1^2 - 2\pi r_1/\F) }
	{ \frac{\pi^2}{\F^2} + (2 k_0 \Delta L n)^2 } +r_1^2
	\\
	T(\Delta L) \approx \frac{ t_1^2 t_2^2 }
	{ \frac{\pi^2}{\F^2} + (2 k_0 \Delta L n)^2 },
	\end{align}
where ${k_0 \approx \omega_l/c}$ is the resonant wavenumber in vacuum, the time-dependent displacement from resonance $\Delta L = v_a(v_b t^2 + t - t_0)$, $v_j$ are fit parameters representing the sweep trajectory, $t_0$ is a fit time offset, and the empty cavity finesse $\F=\pi c \tau/(Ln)$. In these equations, the fit parameters are $\tau$, $\rho_{\mathrm{c}}$, $\rho_{\mathrm{sb}}$, $r_2$, $\epsilon$, $G$ and $G_{\mathrm{slow}}$.

\section{Cavity response to modulated absorption}\label{app:mod_absorption}

Here we use a classical input-output formalism to derive the changes in $V_X$ (the collected and demodulated amplitude quadrature of the Pound-Drever-Hall reflection beat) due to time-varying intra-cavity absorption. We find that $V_X$ responds to modulated absorption as a low pass filter with cutoff frequency equal to the inverse of the cavity's amplitude ringdown time $1/4\pi\tau$, a sensible result since $V_X$ itself is proportional to field.

We start by deriving an equation of motion for the right-moving intra-cavity field $E_{\mathrm{circ}}$ (defined just past the input mirror). At time $t$, this is equal to the incident field transmitted by the front mirror $t_1 E_{\mathrm{in}}(t)$ plus the field $E_{\mathrm{circ}}(t-\Delta t)$ from a roundtrip time ${\Delta t=2Ln/c}$ earlier ($L$ is the cavity length, $n$ the intra-cavity medium refractive index and $c$ the speed of light), after bouncing off both mirrors once and suffering amplitude losses $\delta+\alpha(t)L$:
	\begin{align} 		
	E_{\mathrm{circ}}( t ) &\approx  t_1 E_{\mathrm{in}}(t) + \left(-1 +r_1 + r_2 - \delta - \alpha(t) L\right) E_{\mathrm{circ}} (t - \Delta t)	\label{eq:ecirc_t}
	\end{align}
\noindent
where we again assume lossless mirrors with amplitude reflectivity $r_i$ and transmissivity $t_i$, with $\delta$ capturing their internal losses, and $\alpha(t)$ representing the time-varying extrinsic absorption (power loss per unit length). We have also approximated $\ln (r_i) \approx -1 +r_i$ in the limit $r_i \to 1$. Assuming the roundtrip time is negligible compared to all timescales of interest (i.e., ${2Ln/c\ll \tau}$ and the absorption modulation frequency), in the limit $\Delta t \rightarrow 0$, \cref{eq:ecirc_t} approximates the differential equation
	\begin{align}
    \frac{d E_{\mathrm{circ}}(t) } {d t} &\approx
	- \frac{ \left( 2 - r_1 - r_2 + \delta + \alpha(t) L \right) c E_{\mathrm{circ}} (t ) } {2Ln}
	+ \frac{ t_1 c E_{\mathrm{in}}(t)  } {2Ln}\\ 
	&=
	- \frac{ \kappa_{\mathrm{in}} + \kappa_0 + \kappa_{\alpha} (t)	 } {2} E_{\mathrm{circ}}(t)
	+ \sqrt{ \frac{ \kappa_{\mathrm{in}} c } {2Ln} } E_{\mathrm{in}}(t),
	\label{eq:decircdt}
	\end{align}
in a frame rotating at the cavity resonance frequency $\omega_c$, with total cavity power loss rate 
    \begin{align}
    \kappa(t) &= 
	\frac{c t_1^2 }{ 2 L n } + 
	\frac{c \left(t_2^2 + 2\delta\right) }{ 2 L n } + 
	\frac{c \alpha(t) L }{L n}
	\\
	&\equiv \kappa_{\mathrm{in}}+\kappa_0+\kappa_{\alpha}(t),
	\label{eq:kappa}
    \end{align}
where we approximate ${r_i \approx 1-t_i^2/2}$ in the high-finesse limit ${t_i \to 0}$.
	
We now let the time-varying absorption be a sinusoidal function 
 $\alpha(t) = \alpha_0  \cos(\omega_{\alpha} t)$ (for constants $\alpha_0$ and $\omega_\alpha$), which will in turn modulate the circulating and output fields. Assuming the fluctuations are small, the circulating field can be written as a mean value $\bar{E}_{\mathrm{circ}}$ at the laser frequency plus a small fluctuating term $\Delta E_{\mathrm{circ}}(t)$ due to the modulated absorption, yielding
	\begin{align}\label{eqn:EOM}
	\frac{d	\left( \bar{E}_{\mathrm{circ}}+\Delta E_{\mathrm{circ}}(t) \right)	} {dt} &=
	-\frac{\kappa_{\mathrm{in}}+\kappa_0+\kappa_{\alpha}(t)}{2} 	
	\left( \bar{E}_{\mathrm{circ}}+\Delta E_{\mathrm{circ}}(t) \right)
	+\sqrt{ \frac{\kappa_{\mathrm{in}}c}{2Ln} } \bar{E}_{\mathrm{in}}.
	\end{align}
Keeping only the time-varying terms, and neglecting terms quadratic in the fluctuation, we find:
	\begin{align}
	\frac{d	\left(\Delta E_{\mathrm{circ}}(t) \right)} {dt} 
	&\approx
	- \frac{ \kappa_{\mathrm{in}} + \kappa_0 } {2} \Delta E_{\mathrm{circ}}(t)
	- \frac{ \kappa_{\alpha} (t) } {2} \bar{E}_{\mathrm{circ}}.
	\end{align}	
Substituting $\kappa_{\alpha}(t)=c \alpha_0 \cos(\omega_{\alpha} t)/n $ and working in the Fourier domain\footnotemark,
	\begin{align}
	\Delta E_{\mathrm{circ}}(\omega) &=
	-\underbrace{\frac{1}{ -i\omega+ \frac{ \kappa_{\mathrm{in}} + \kappa_0 } {2} }}_{\chi(\omega)}
	\Big( \delta(\omega-\omega_{\alpha}) + \delta(\omega+\omega_{\alpha}) \Big)	
	\frac{ c\alpha_0 }{4n} \bar{E}_{\mathrm{circ}}	
	\label{eq:deltaEcirc1}
	\end{align}	
where $\delta(\omega\pm\omega_{\alpha})$ is the Dirac delta function, the Fourier frequency $\omega$ is the detuning from resonance, and we defined the cavity susceptibility $\chi(\omega) \equiv 1/\left( -i \omega + \frac{ \kappa_{\mathrm{in}} + \kappa_0 } {2} \right)$.

\footnotetext{We use the following convention for the Fourier transform of a function $f(t)$:
$$ FT \{f(t)\} = f(\omega) =\frac{1}{2\pi} \int_{-\infty}^{+\infty} f(t) e^{i \omega t} dt \ .$$
}

\begin{figure}[ht]
    \centering
    \includegraphics[width=.9\textwidth,trim=0 0 0 0,clip]
    {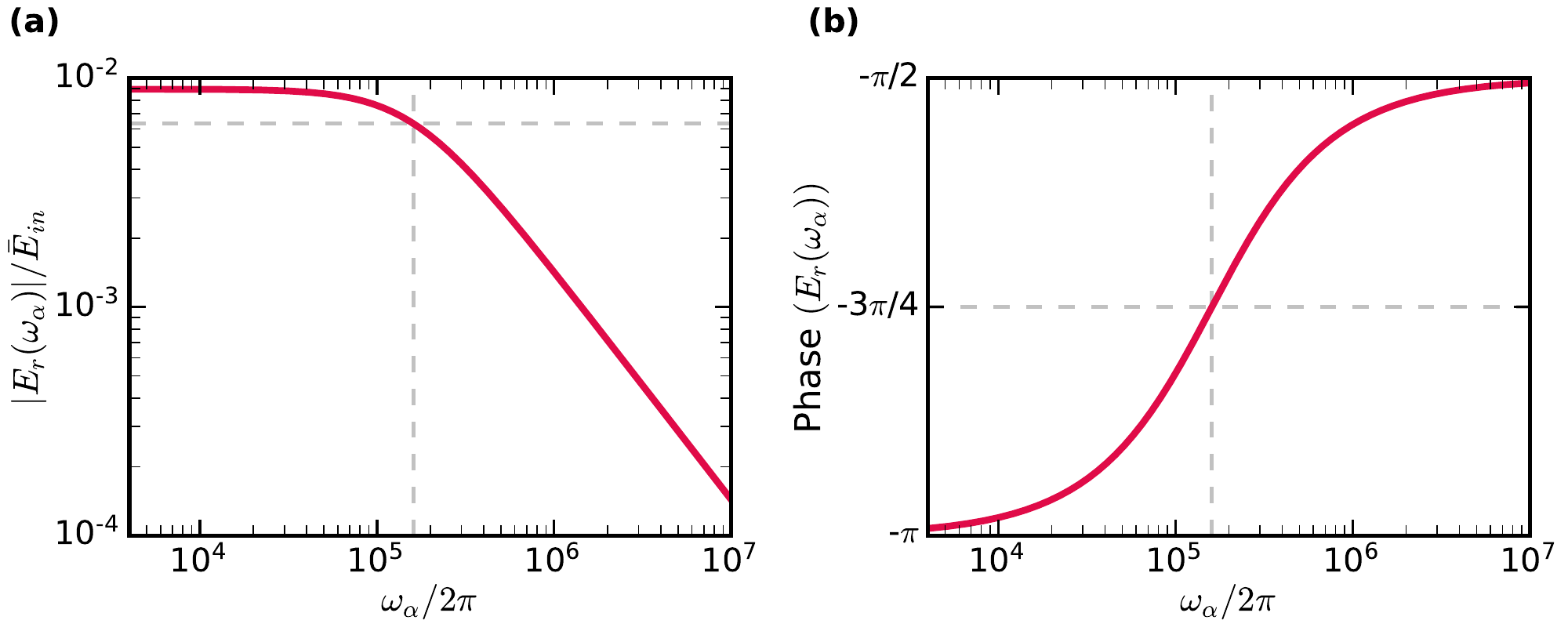}
    \caption{\textbf{Modulated intra-cavity absorption.} 
    (a)~Magnitude and (b)~phase of the outgoing absorption modulation sideband $E_r(\omega_{\alpha})$ (\cref{eq:Er_omega_a}) as a function of the absorption modulation frequency $\omega_{\alpha}/2\pi$. We observe the cavity low-pass behaviour with cutoff frequency $1/4\pi\tau=160$ kHz (vertical dashed lines). Horizontal dashed lines mark the -3 dB point in magnitude and $3\pi/4$ phase-shift.
    }
    \label{fig:alpha_mod}
\end{figure}

Meanwhile, $\bar{E}_{\mathrm{circ}}$ can be found from the time-independent terms of \cref{eqn:EOM}:
	\begin{align}
	\bar{E}_{\mathrm{circ}} 
	&= \underbrace{\frac{2}{\kappa_{\mathrm{in}} + \kappa_0}}_{\chi(0)} \sqrt{ \frac{\kappa_{\mathrm{in}} c }{ 2 Ln } } \bar{E}_{\mathrm{in}},		\label{eq:ecirc}
	\end{align}
which, upon substitution into 
\cref{eq:deltaEcirc1}, yields the modulation in the circulating power:
	\begin{align}	\label{eq:deltaEcirc_w}
	\Delta E_{\mathrm{circ}}( \omega) &= -\chi(\omega) \Big( \delta(\omega-\omega_{\alpha}) + \delta(\omega+\omega_{\alpha}) \Big)	
	\frac{ c\alpha_0 }{4n}	\chi(0) \sqrt{ \frac{\kappa_{\mathrm{in}} c }{ 2 Ln } } \bar{E}_{\mathrm{in}}.
	\end{align}	
The cavity reflected field, given by $E_{r}(t)= -r_1 \bar{E}_{\mathrm{in}} + t_1 \left( \bar{E}_{\mathrm{circ}} +\Delta E_{\mathrm{circ}}(t)\right),$ in the frequency domain becomes 
	\begin{align}
	E_{r}(\omega) =& -r_1 \bar{E}_{\mathrm{in}} + t_1 \left( \bar{E}_{\mathrm{circ}} +\Delta E_{\mathrm{circ}}(\omega)\right).
	\end{align}
Substituting \cref{eq:ecirc,eq:deltaEcirc_w} and $t_1 = \sqrt{2 \kappa_{\mathrm{in}} Ln/c }$ (\cref{eq:kappa}) into this, we find that there are only three frequencies with nonzero field, at DC and $\pm \omega_a$:
	\begin{align}
	E_{r}(0) =&\left( -r_1 + \kappa_{\mathrm{in}} \chi(0) \right)	\bar{E}_{\mathrm{in}}	 
 	\\
	E_r(\pm\omega_\alpha) =& - \kappa_{\mathrm{in}} \chi(0) \chi(\pm\omega_{\alpha})
	 	\frac{ c \alpha_0 }{ 4n } \bar{E}_{\mathrm{in}}.   	\label{eq:Er_omega_a}
	\end{align}
So, recalling that we are working in a frame rotating at $\omega_c$, the out-going reflected field will have a carrier at the resonance frequency $\omega_c$ and two sidebands detuned by the absorber frequency at $\omega_c \pm\omega_{\alpha}$. 
On \cref{fig:alpha_mod}(a) and (b), we plot the magnitude and phase of ${E_{r}(\omega_{\alpha})}$ (\cref{eq:Er_omega_a}), and we can observe that the cavity acts as a low-pass filter with cutoff frequency ${\omega_{\text{cutoff}}= (\kappa_{\mathrm{in}}+\kappa_0)/2 \equiv 1/2\tau}$ for intra-cavity modulated absorption, as would be expected from the cavity amplitude decay time.

To get an expression for the time-modulated $V_X(t)$, we have to account for the phase-modulation of the incoming laser beam, i.e., include the promptly reflected, off-resonant carrier and (upper) sideband, and lock the other (lower) sideband on resonance with the cavity ($\omega_c\to\omega_l-\Omega$, where $\omega_l$ is the carrier frequency and $\Omega$ is the modulation frequency).
The incoming field can be written as 
    \begin{align}
    E_{\mathrm{in,PDH}}(t)&= E_0 e^{i\omega_l t}\sqrt{\rho_{\mathrm{c}}} + E_0 e^{i (\omega_l+\Omega)t}\sqrt{\rho_{\mathrm{sb}}} - E_0 e^{i (\omega_l-\Omega)t}\sqrt{\rho_{\mathrm{sb}}}
    \end{align}
where now we let the incoming field be a fast-varying function of time with amplitude $E_0$.
Upon reflection from the cavity, having the lower sideband locked on resonance and the intra-cavity field time-modulated by transient absorption, the outbound field becomes
    \begin{align}   \label{eq:Erpdh_t}
	E_{r, \mathrm{PDH}}(t) 
	=& - r_1 E_0\left( e^{i\omega_l t}\sqrt{\rho_{\mathrm{c}}} + e^{i (\omega_l+\Omega) t}\sqrt{\rho_{\mathrm{sb}}}\right)
	\nonumber\\
	&-\Big(-r_1 E_0 + t_1 \left( \bar{E}_{\mathrm{circ}} +\Delta E_{\mathrm{circ}}(t) \right) \Big) e^{i(\omega_l-\Omega) t} \sqrt{\rho_{\mathrm{sb}}} . 
	\end{align}    
We can find $\Delta E_{\mathrm{circ}}(t)$ by taking the inverse Fourier transform of \cref{eq:deltaEcirc_w}:
    \begin{align}   \label{eq:deltaEcirc_t}
    \Delta E_{\mathrm{circ}}(t) =& \;
     - \Big( e^{i\omega_{\alpha}t} \chi(-\omega_{\alpha}) + e^{-i\omega_{\alpha}t} \chi(\omega_{\alpha}) \Big)
    \frac{c\alpha_0}{4n}\chi(0) \sqrt{\frac{\kappa_{\mathrm{in}}c}{2Ln}} E_0.
    \end{align}
Substituting \cref{eq:deltaEcirc_t,eq:ecirc} into \cref{eq:Erpdh_t}, noting that $-r_1+\kappa_{\mathrm{in}}\chi(0)=r_{\mathrm{res}}$,
    \begin{multline}
    	E_{r, \mathrm{PDH}}(t) 
	= - r_1 E_0 \left( e^{i\omega_l t}\sqrt{\rho_{\mathrm{c}}} + e^{i(\omega_l+\Omega)t}\sqrt{\rho_{\mathrm{sb}}} \right)- r_{\mathrm{res}}E_0e^{i(\omega_l-\Omega)t}\sqrt{\rho_{\mathrm{sb}}}
	\\
    +(r_1+r_{\mathrm{res}})E_0\sqrt{\rho_{\mathrm{sb}}} \frac{c \alpha_0}{4n}
    \left( e^{i(\omega_l-\Omega+\omega_{\alpha})t} \chi(-\omega_{\alpha}) +  e^{i(\omega_l-\Omega-\omega_{\alpha})t} \chi(\omega_{\alpha}) \right).
    \end{multline}
We finally find $V_X(t)$ by first calculating $P_{R,\mathrm{PDH}}(t)=E_{r, \mathrm{PDH}}(t)^* E_{r, \mathrm{PDH}}(t)$, collecting it with a photodetector, shifting by $\pi/2$ and mixing with a reference at $\sin(\Omega t)$. Neglecting terms oscillating at $\geq\Omega$, 
    \begin{multline}    \label{eq:Vx_t}
    V_X(t) = - 2 G P_\mathrm{in} \sqrt{\rho_{\mathrm{c}}\rho_{\mathrm{sb}}} \; r_1 \left(r_1+r_{\mathrm{res}}\right) \left(1 -
    \Big( \chi(-\omega_{\alpha})+\chi(\omega_{\alpha}) \Big) \frac{ c\alpha_0 }{4n} \cos(\omega_{\alpha}t) \right.
    \\
    \left. - \Big( \chi(-\omega_{\alpha})-\chi(\omega_{\alpha}) \Big) \frac{ c\alpha_0 }{4n} i\sin(\omega_{\alpha}t)
    \right)
    \end{multline}
where $G$ is the gain of the detector and mixer circuits and we define $P_\mathrm{in}\equiv E_0^2$. We conclude that intra-cavity time-modulated absorption adds a fluctuating term at the absorption modulation frequency $\omega_{\alpha}$ to $V_X$, which is filtered by the cavity behaving as a low-pass with cutoff frequency $1/4\pi\tau$, determined by the cavity susceptibility $\chi(\pm\omega_{\alpha})= 1/\left(\frac{1}{2\tau}\mp i\omega_{\alpha}\right)$.

\section{Subtraction of classical noise}\label{app:sub}

To better understand and improve technical noise, we simultaneously monitor the error signal, VCO output power, and incident laser power, which provide some information about phase noise, changes in modulation frequency, and laser amplitude noise, respectively. If one of these quantities is strongly correlated with our excess noise, we can appropriately scale, delay, filter, and subtract these signals from $V_X$ in the time domain, thereby helping identify the most important noise sources and partially mitigating their impact. When possible, we determine the parameters for subtraction from our theoretical model of the system. The rest are the result of minimization of the average PSD after time-domain subtraction over a targeted frequency range. All subtraction parameters are estimated from a signal-free ``calibration'' data set, and fixed for the subtraction process performed on the ``measurement'' data set.

Fluctuations in the intensity of the incoming laser produce correlated noise in $V_X$; for small fluctuations, we expect a frequency-dependent, linear scaling $H(\omega)$ in the frequency domain, such that $P_X(\omega) \propto H(\omega) P_\mathrm{in}(\omega)$.
We derived $H(\omega)$ in a manner similar to the derivation presented in Appendix~\ref{app:mod_absorption}, considering sinusoidal modulation of the incoming beam's intensity at frequency $\omega$ and calculating the appropriate quadrature of the PDH signal.  
The resulting transfer function for a cavity with a sideband on resonance is
\begin{align} 
  H(\omega)
    =&  \sqrt{\rho_{\mathrm{c}}\rho_{\mathrm{sb}}} r_1^2 + \frac{1}{4}\sqrt{\rho_{\mathrm{c}}\rho_{\mathrm{sb}}}r_1\Big(r(-\omega)+r(0) + r^*(\omega)+r^*(0)\Big),\\
    \approx&  \frac{1}{2} \sqrt{\rho_{\mathrm{c}}\rho_{\mathrm{sb}}} t_1^2 
    \Big( \frac{\F}{\pi}+ \frac{1} { \frac{\pi}{\F} + 2iLn\omega/c} \Big),
\end{align}
using the high finesse cavity fractional amplitude reflection $r(\omega) \approx \frac{ t_1^2 } { \pi/\F - 2iLn\omega/c } - r_1$. This filter is applied to the pick-off photodiode voltage by multiplying in the Fourier domain then transforming back to the time domain (though this can also be done in real time), which produces a filtered intensity voltage $V_{FI}(t)$ that should be on equal footing with $V_X$. 

In principle, fluctuations in the VCO output power and fluctuations in the error monitor (a proxy for cavity detuning) also have a frequency-dependent influence on $V_X$. But we found the results below to be insensitive to changes in the subtraction method, likely due to noise in the auxiliary measurements we wish to subtract.

Since the signals are measured with different devices having different gains and delays at different locations in the beam path, subtraction requires that we scale each voltage by a constant and find the appropriate delay. In the case of the filtered laser intensity and VCO amplitude, we can use the mean value of the signals to determine the scaling factor because
\begin{align}
    V_X(t) &= \bar{V}_X ( 1 + n_L(t))(1 + n_O(t)),\\
    &\approx \bar{V}_X ( 1 + n_L(t) + n_O(t))
\end{align}
is approximately proprotional to both, where $n_L(t)$ is the relative classical noise in the laser power and $n_O(t)$ is the relative fluctuation in VCO output amplitude. Specifically, 
\begin{align}
    n_L(t) &= \frac{V_{FI}(t)}{\bar{V}_{FI}}-1 \\
    n_O(t) &\approx \frac{1}{2}\left(\frac{V_{OP}(t)}{\bar{V}_{OP}} -1\right),
\end{align}
where mean values are denoted with a bar and $V_{OP}(t)$ is a voltage proportional to the power of the VCO output (note that, for small fluctuations, the noise in the VCO amplitude is half the noise in its power). Ultimately we obtain a noise-reduced subtracted voltage $V_{\mathrm{sub}}$ via
\begin{align}
    V_{\mathrm{sub}}(t) = V_X(t) - \bar{V}_X\left(\frac{V_{FI}(t - \Delta t_{FI})}{\bar{V}_{FI}}-1\right) - \frac{\bar{V}_X}{2}\left(\frac{V_{OP}(t - \Delta t_{OP})}{\bar{V}_{OP}} -1\right) - a_{E} V_{E}(t-\Delta t_{E}),
\end{align}
where we have now included measurements of the error monitor voltage $V_{E}$ with unknown scaling factor $a_{E}$, and added empirical time delays $\Delta t_{FI}$, $\Delta t_{OP},$ and $\Delta t_{E}$. 

\begin{figure}
    \centering
    \includegraphics[width=0.7\textwidth]{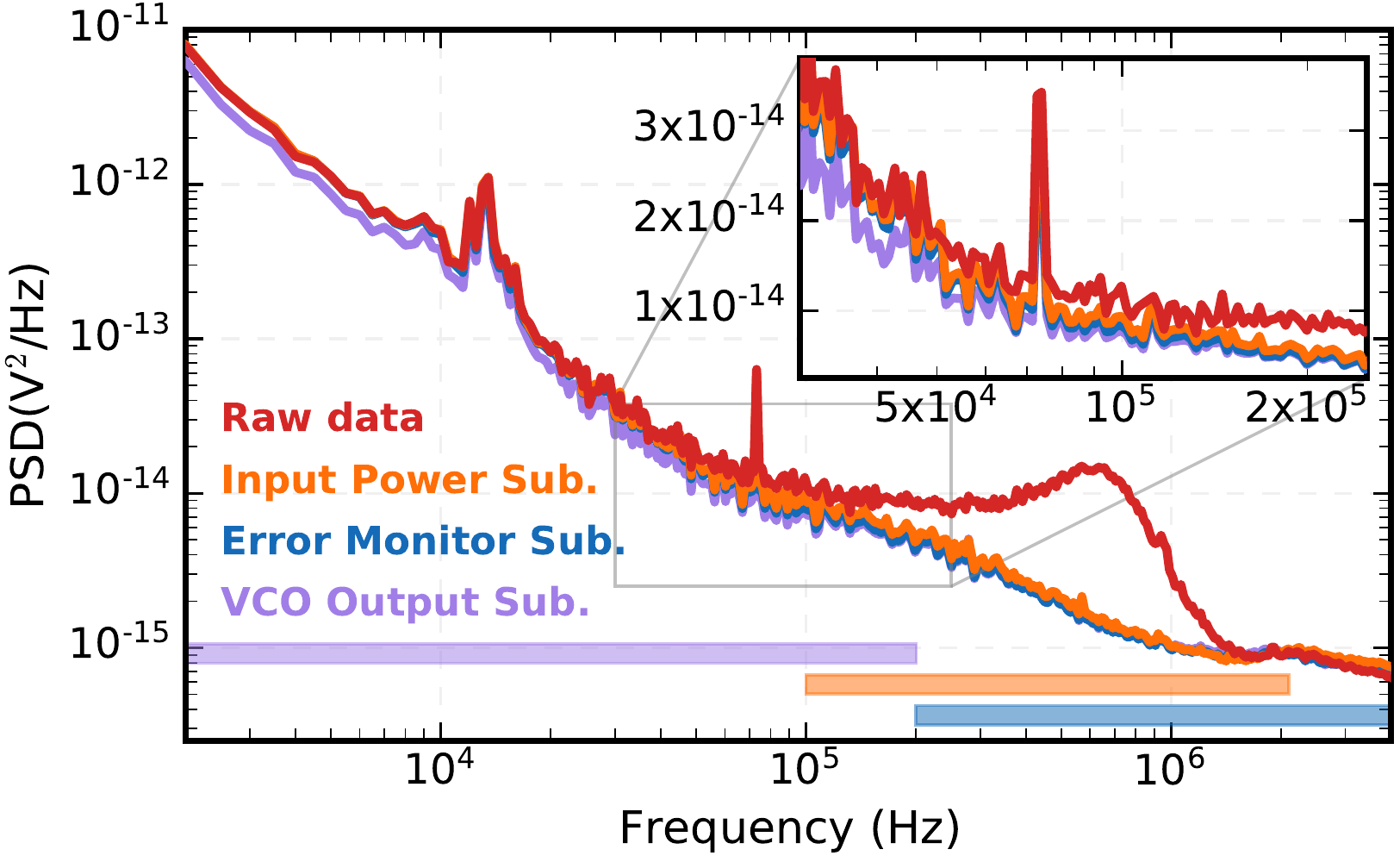}
    \caption{\textbf{Noise reduction by time-domain subtraction.} The red curve shows the PSD of locked $V_X$, while orange, blue, and purple show the signal PSD after successively subtracting the appropriately scaled and shifted input power monitor, error monitor, and VCO output. The input power is filtered with the appropriate transfer function, derived in the text, as well as delayed by $\Delta t_{FI}=50$ ns (likely not a real delay, and instead an empirical correction that partially compensates for differences in the equipment used to measure the transfer function). Time delays for the other subtractions were not found to improve the result and were left at zero.  Horizontal bars show the bandwidth targeted by each subtraction.}
    \label{fig:subtraction}
\end{figure}

To determine the scaling factor for the error signal and the appropriate time delays for each noise source, we resort to empirical methods, minimizing the area under the PSD of $V_{\mathrm{sub}}$ over a different frequency range for each signal (chosen based on where noise is observed to be most correlated). Figure 6 shows the results of of subtracting first the input power, then the error monitor, then VCO noise, illustrating that (at least with our equipment) classical laser noise subtraction yields the greatest benefit for improving the sensitivity of our system. It is worth noting that this data set was taken with the cavity precisely tuned to sideband resonance; any static detuning (arising, for example, from drifts in the lock setpoint) makes the system highly sensitive to detuning noise, in which case we found that subtraction of the error monitor voltage largely cancels the additional noise. 

\end{appendices}


\begin{backmatter}
\bmsection{Funding}
This research was conducted as part of the activities of MEDTEQ$+$, thanks to the financial support of the Ministry of Economy and Innovation - PSOv2b program, cofounded by TransMedTech (Canada First Research Excellence Fund), Mitacs, the MUHC Foundation and Varian Medical Systems.  L. Childress acknowledges relevant support from the National Science and Engineering Research Council (NSERC RGPIN 435554-13,  RGPIN-2020-04095), Canada Research Chairs (229003 and 231949), and the Canada Foundation for Innovation (Innovation Fund 2015 project 33488 and LOF/CRC 229003); L. Childress is a CIFAR fellow in the Quantum Information Science program. J. Sankey acknowledges relevant support from the Natural Sciences and Engineering Research Council of Canada (NSERC RGPIN 2018-05635), Canada Foundation for Innovation (LOF/CRC 228130, Innovation 36423), Canada Research Chairs (235060), Institut Transdisciplinaire d'Information Quantique (INTRIQ), and the Centre for the Physics of Materials (CPM) at McGill.

\bmsection{Disclosures}
The authors declare no conflicts of interest.

\bmsection{Data Availability}
Data underlying the results presented in this paper are not publicly available at this time but may
be obtained from the authors upon reasonable request.

\end{backmatter}

\bibliography{CavitySensing}

\end{document}